\documentclass[twocolumn]{aastex631}
\usepackage{amsmath}
\usepackage{colortbl}
\usepackage{multirow}
\usepackage{subfigure}
\usepackage{graphicx}
\usepackage{xspace}
\usepackage{xcolor}
\usepackage{ulem}
\usepackage{soul}
\newcommand {\chandra}{\textit{Chandra}\xspace}

\newcommand {\arcm}{{$^{\prime}$}\xspace}
\begin{document}

\title{Comparison between the emission torus and the measured toroidal magnetic field for the Crab and Vela nebula}

\author[0000-0002-9370-4079]{Wei Deng}
\affiliation{Guangxi Key Laboratory for Relativistic Astrophysics, School of Physical Science and Technology, Guangxi University, Nanning 530004, China}

\author[0000-0002-0105-5826]{Fei Xie}
\correspondingauthor{Fei Xie}
\email{xief@gxu.edu.cn}
\affiliation{Guangxi Key Laboratory for Relativistic Astrophysics, School of Physical Science and Technology, Guangxi University, Nanning 530004, China}
\affiliation{INAF Istituto di Astrofisica e Planetologia Spaziali, Via del Fosso del Cavaliere 100, 00133 Roma, Italy}

\author[0009-0007-8686-9012]{Kuan Liu}
\affiliation{Guangxi Key Laboratory for Relativistic Astrophysics, School of Physical Science and Technology, Guangxi University, Nanning 530004, China}

\author[0000-0002-3776-4536]{Mingyu Ge}
\affiliation{Key Laboratory of Particle Astrophysics, Institute of High Energy Physics, Chinese Academy of Sciences, Beijing 100049, China}

\author[0000-0003-3127-0110]{Youli Tuo}
\affiliation{Institut f\"{u}r Astronomie und Astrophysik, Kepler Center for Astro and Particle Physics, Eberhard Karls Universit\"{a}t T\"{u}bingen, Sand 1, 72076 T\"{u}bingen, Germany}

\author[0000-0001-8916-4156]{Fabio La Monaca}
\affiliation{INAF Istituto di Astrofisica e Planetologia Spaziali, Via del Fosso del Cavaliere 100, 00133 Roma, Italy}
\affiliation{Dipartimento di Fisica, Universit$\grave{a}$ degli Studi di Roma “Tor Vergata”, Via della Ricerca Scientifica 1, 00133 Roma, Italy}
\affiliation{Dipartimento di Fisica, Universit$\grave{a}$ degli Studi di Roma “La Sapienza”, Piazzale Aldo Moro 5, 00185 Roma, Italy}

\author[0000-0003-0331-3259]{Alessandro Di Marco}
\affiliation{INAF Istituto di Astrofisica e Planetologia Spaziali, Via del Fosso del Cavaliere 100, 00133 Roma, Italy}

\author[0000-0002-7044-733X]{En-wei Liang}
\affiliation{Guangxi Key Laboratory for Relativistic Astrophysics, School of Physical Science and Technology, Guangxi University, Nanning 530004, China}

\begin{abstract}
Polarization measurements provide insight into the magnetic field, a critical aspect of the dynamics and emission properties around the compact object. In this paper, we present the polarized magnetic field of the Crab outer torus and the Vela arc utilizing Imaging X-ray Polarimetry Explorer observation data. The polarization angle (PA) measured for the Crab outer torus exhibits two monotonic evolutions along the azimuth angle, which are consistent with the normal line of the elliptical ring. There is a slight increase in PA along the azimuth angle for both the inner arc and the outer arc of the Vela nebula. The polarized magnetic vector along the outer torus of the Crab nebula shows the polarized magnetic field aligns with Crab outer torus structure. The PA variation along the Crab outer torus suggests a bulk flow speed of 0.8$c$. Meanwhile, the Vela nebula polarized magnetic field does not exactly align with the Vela arc structure. We noted that the Crab nebula possesses a polarized toroidal magnetic field, where as the Vela nebula exhibits an incomplete toroidal magnetic field.
\end{abstract}


\section{Introduction} 
\label{sec:intro}
The nature of radiation mechanism of high-energy particles in the magnetized plasma of the pulsar wind nebulae (PWNe) is a long-standing problem. Relativistic particles in PWNe produce emission observed from the radio to $\gamma$-ray band. The processes responsible for particle acceleration and radiation across the electromagnetic spectrum remain unclear. Theoretical models suggest that synchrotron radiation dominates at radio, optical, and X-ray, while inverse Compton scattering contribute to high-energy emissions \citep{2006ARA&A..44...17G,2014RPPh...77f6901B}. However, the exact nature of these mechanisms and the factors influencing their efficiency are still under debate \citep{2003MNRAS.344L..93K,2009ApJ...698.1523S,2017ApJ...847...57Z}.

High-resolution \chandra images allow us to study their prominent features in the X-ray band close to the center region. Both Crab PWN \citep{2000ApJ...536L..81W} and Vela PWN \citep{2001ApJ...556..380H} exhibit features characterized by ``torus-jet''. It is now known that compact X-ray structures are formed by the interaction between the anisotropic pulsar wind and the surrounding medium in the toroidal magnetic field. Relativistic magnetohydrodynamic models of PWNe indicate that the pulsar wind outflow is anisotropic, which is more energetic in the equatorial plane than along the polar axis \citep{2004A&A...421.1063D, 2004MNRAS.349..779K, 2006A&A...453..621D}.

\begin{figure*}[ht]
\centering
\subfigure{\includegraphics[width=1.0\columnwidth]{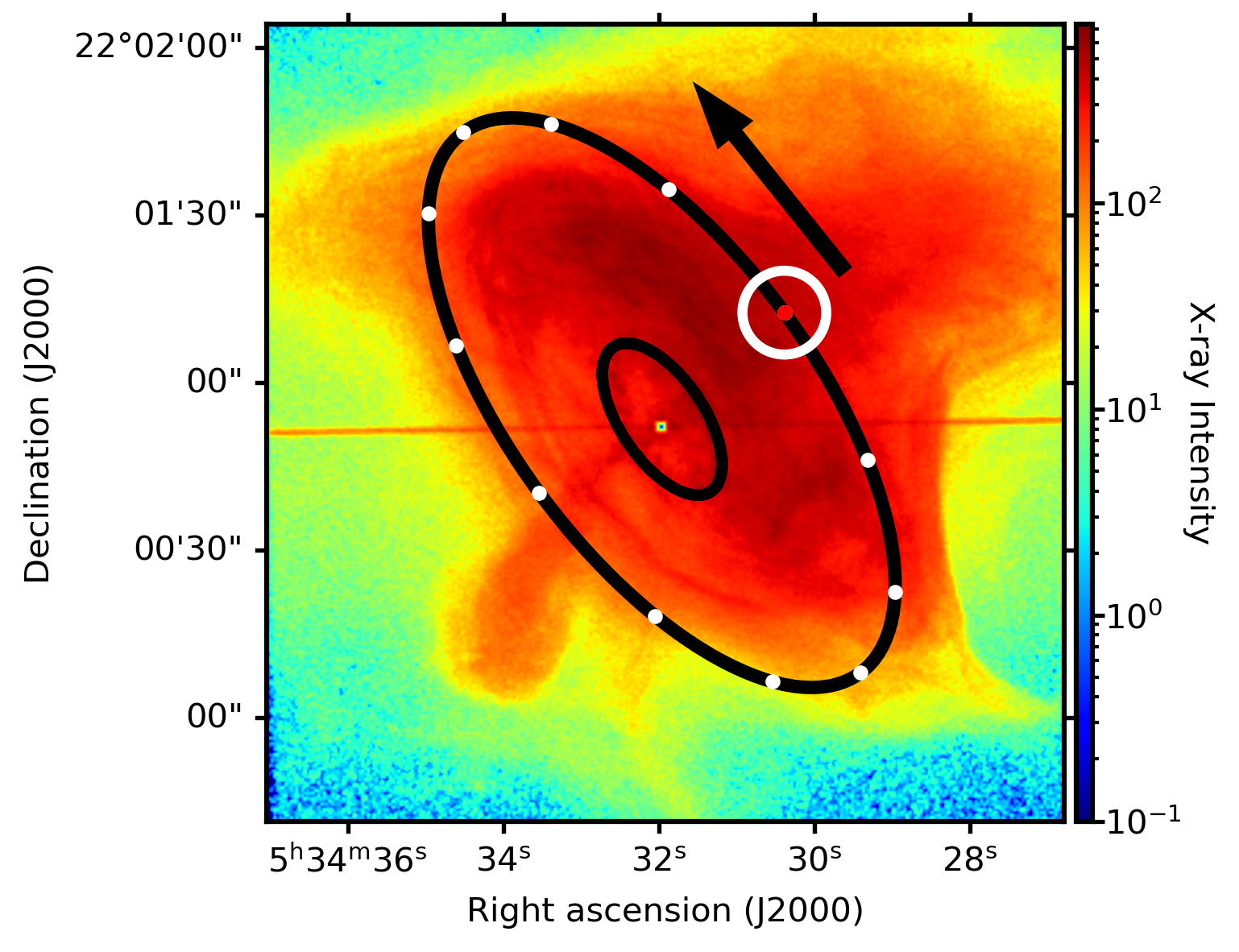}}
\subfigure{\includegraphics[width=1.0\columnwidth]{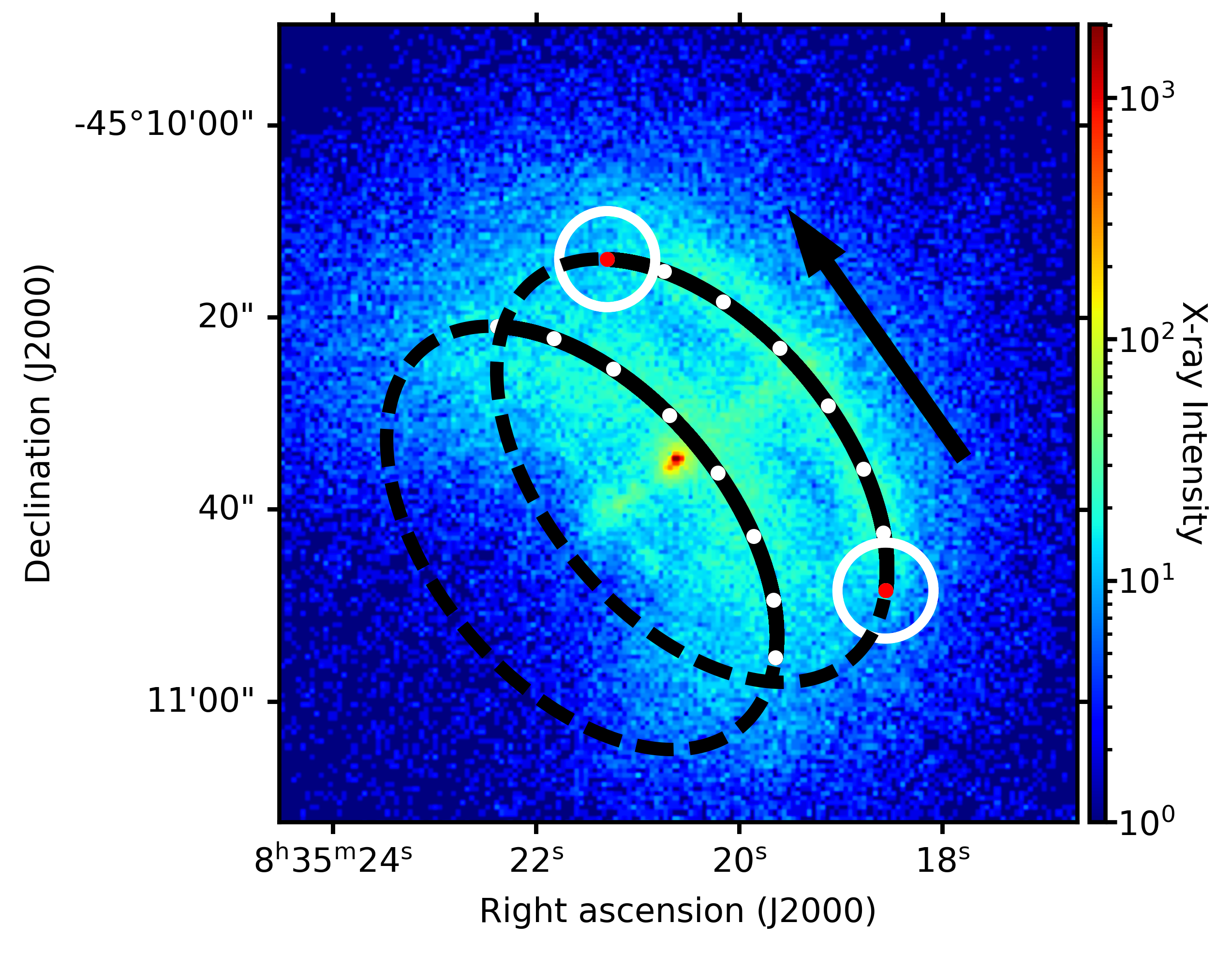}}

\caption{Image of the Crab nebula (left) and Vela nebula (right) observed by \chandra. The Crab nebula exhibits a fully formed double-ring torus-jet feature, while the Vela nebula presents an incomplete double-arcs torus-jet characteristic. The black inclined ellipses are the geometric models proposed in \cite{2004ApJ...601..479N} for Crab and Vela emission torus. The solid elliptical arcs in the right panel represent Vela emission double-arcs. The white points represent the centers of the circular regions for analysis, with the size marked by the white circles, and the centers are in red dots. Arrows are for the region-select direction.}
\label{fig:image_twonebula}
\end{figure*}

Crab pulsar and its nebula, hosted in the center of supernova SN1054 \citep{1978A&A....70..419W}. Its most prominent X-ray features are the inner ring, the outer torus, and the jet along the torus axis \citep{2000ApJ...536L..81W}, as shown in the left panel of Figure \ref{fig:image_twonebula}. The inner ring is usually associated with the termination shock, which has an elliptical shape with a semi-major axis of 15.6$^{\prime\prime}$ (0.15 pc) and a semi-minor axis of 7.5$^{\prime\prime}$ (0.07 pc), assuming a distance of 2 kpc \citep{2004ApJ...601..479N}. The outer torus displays circumferential fibrous structures, with surface brightness varying considerably along the toroidal azimuth. The semi-major and semi-major axes of the boundary of the outer torus are approximately 60$^{\prime\prime}$ (0.58 pc) and 27.2$^{\prime\prime}$ (0.26 pc), respectively. 

Vela PWN, powered by PSR B0833-54 \citep{1968Natur.220..340L}, is a complex and variable astronomical system. It consists of two non-thermal toroidal arcs, an inner jet and a curved outer jet, shown in the right panel of Figure \ref{fig:image_twonebula}. The two arcs are roughly elliptical in shape and have a similar curvature \citep{2001ApJ...556..380H}. The 10$^{\prime\prime}$ (0.014\,pc) inner jet, and the curved outer jet extending to 100$^{\prime\prime}$ (0.14\,pc) and exhibiting significant variability in both brightness and shape, are approximately perpendicular to the double-arcs. These physical sizes are calculated on the basis of the distance to the Vela nebula of 290\,pc \citep{2001ApJ...561..930C}.

It is commonly accepted that PWNe have a toroidal magnetic field in which relativistic particles radiate emissions, as confirmed by multiple simulations \citep{2003MNRAS.344L..93K, 2004A&A...421.1063D}. 

The spatially-resolved polarization map obtained by the Cambridge One-Mile telescope displays a complex magnetic field structure over the Crab PWN at 2.7 GHz and 5 GHz \citep{1972MNRAS.157..229W}. Optical observations by Hubble Space Telescope reveal that Crab nebula has a synchrotron knot and wisps with polarization degree (PD) of 59\% and 40\% respectively \citep{2013MNRAS.433.2564M}. The Vela nebula exhibits a toroidal magnetic field with average PD reaching 60\% in the radio band \citep{2002ASPC..271..187B, 2003MNRAS.343..116D}, and no corresponding optical emission is detected \citep{2003ApJ...594..419M}. X-ray polarimetry could allow us to investigate the magnetic field near the site of particle acceleration. Although X-ray polarimetry measurements were conducted on the Crab pulsar and nebula together \citep{1978ApJ...220L.117W,2017NatSR...7.7816C,2018NatAs...2...50V,2020NatAs...4..808F}, a spatially-resolved polarimetric analysis in this band was not possible until the successful launch of Imaging X-ray Polarimetry Explorer (IXPE).

IXPE observations reveal a toroidal magnetic field with an average PD of 19\% for the nebula and an asymmetric PD distribution in the 2--8\,keV energy band \citep{2023NatAs...7..602B}. In the case of the Vela PWN, X-ray polarization measurements show a toroidal magnetic field with a high image-averaged PD of 45\%, and PD$>$60\% manifested in some local regions, approaching the synchrotron limit \citep{2022Natur.612..658X, 2023ApJ...959L...2L}. This finding implies that Vela PWN has a highly ordered magnetic field in which relativistic particles emit synchrotron radiation.

Using the high spatial resolution image from \textit{Chandra} and the spatial polarimetry data from IXPE, we can investigate the relationship between polarization properties and the detail emission features of PWNe. In this work, we utilize data obtained from the IXPE to enhance our understanding of the polarized magnetic field properties within the Crab outer torus and Vela arcs. We present the X-ray polarization results of the regions along the Crab outer torus and Vela arcs. This paper is organized as follows: Section \ref{sec:Data} describes the IXPE observations and data reduction. In Section \ref{sec:results}, we provide a detailed description of the data analysis procedures used to extract the polarization properties along the Crab torus and Vela arcs. We summarize the results in Section\,\ref{sec:discussion}. 

\begin{figure*}[ht]
\centering
\subfigure{\includegraphics[width=1.0\columnwidth]{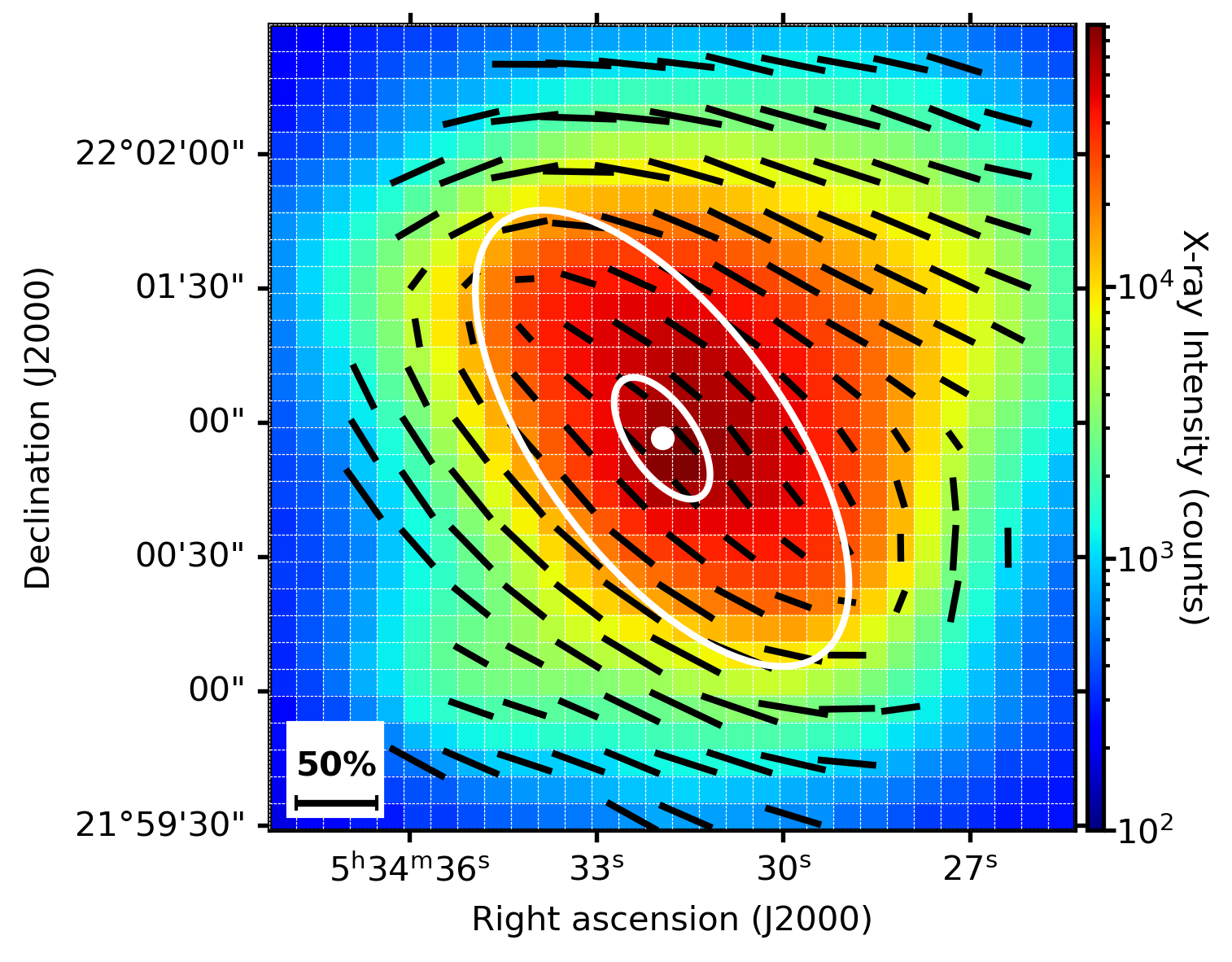}}
\subfigure{\includegraphics[width=1.0\columnwidth]{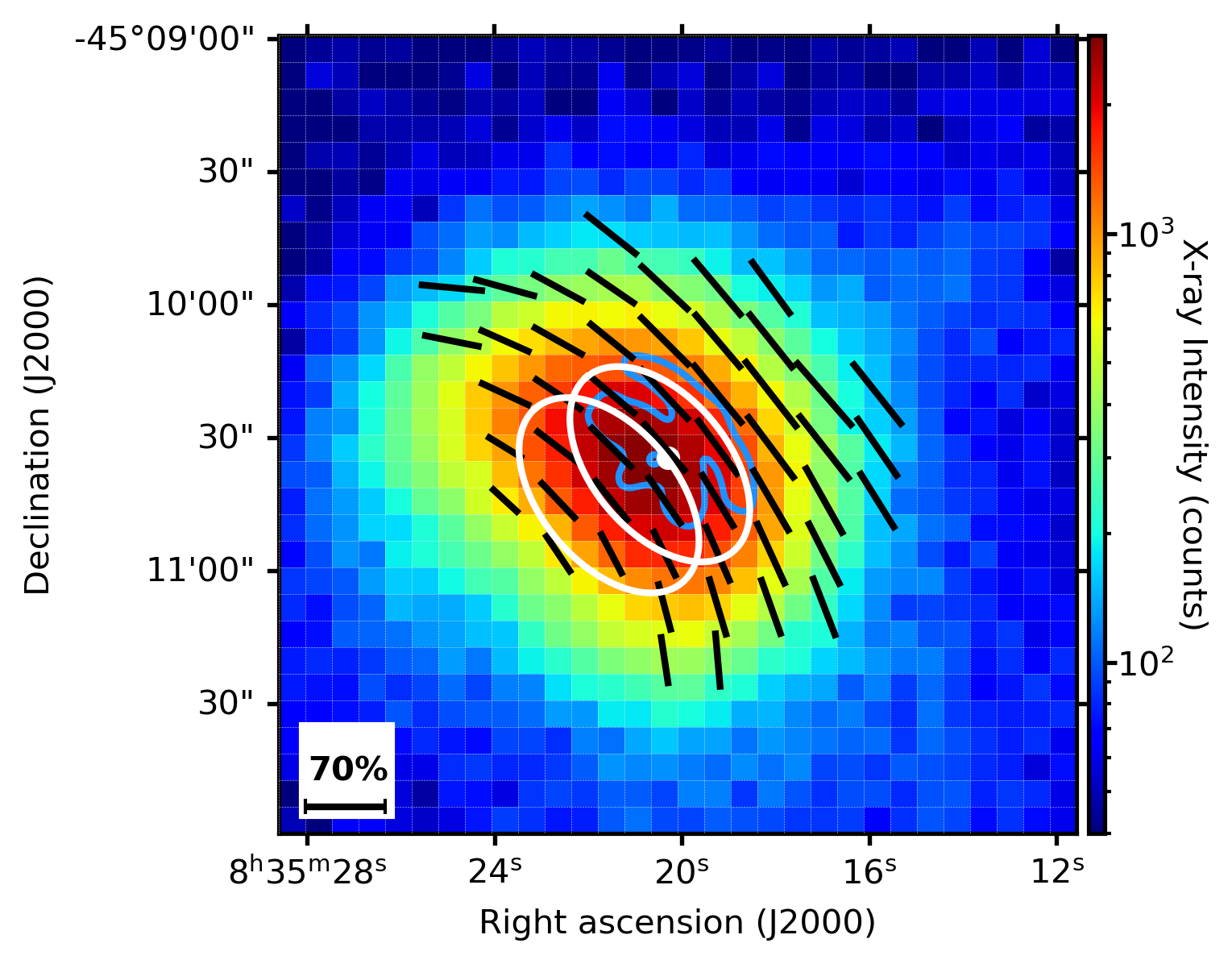}}
\caption{The intensity map of the Crab PWN (left) and the Vela PWN (right) in the 2--8 keV energy band, overlaid with the reconstructed polarization vectors (magnetic field, in black lines) cut at the $5\sigma$ significance level.
The pulsars are located in the center, shown in white dots. The white inclined ellipses are the geometric models proposed in \cite{2004ApJ...601..479N} for Crab and Vela emission torus. The light-blue contours are obtained from the \textit{Chandra} observations for Vela PWN. 
The black bar on the bottom left shows the maximum measured X-ray PD.
The frame sizes are 3\arcm $\times$ 3\arcm.}
\label{fig:map}
\end{figure*}

\section{Observation and Data Reduction}
\label{sec:Data}
IXPE \citep{2021AJ....162..208S, 2022JATIS...8b6002W} was launched on December 9, 2021, and it is the first dedicated X-ray polarimeter with imaging capability so far. It consists of three co-aligned X-ray telescopes, each equipped with an imaging photoelectric polarimeter based on a gas pixel detector \citep{Costa2001, 2021APh...13302628B}. The equipped Wolter-1 mirror module assembly enables a field of view of $12.9^{\prime}\times12.9^{\prime}$, and an angular resolution of $\le 30^{\prime\prime}$ in half-power diameter \citep{2022JATIS...8b6002W}. This allows spatially resolved polarimetry of the extended PWNe for the first time. During more than two years of operation, it has provided serendipitous results thanks to its advanced capabilities for measurement in imaging, photometry, spectroscopy, and polarimetry for many sources.

Crab and Vela were the first two PWNe observed by IXPE in 2022, with results reported in \cite{2023NatAs...7..602B,2024arXiv240712779W} and \cite{2022Natur.612..658X} respectively. In this work, we use the same data of these publications through HEASARC (OBSID 01001099, 02002099, and 02006001 for the Crab and OBSID 01001299 for the Vela). 
\cite{2023NatAs...7..602B} reported the results using the first observation of the Crab PWN, and \cite{2024arXiv240712779W} used all the three observations, with a total exposure time of approximately 300 ks. In this work, we applied the same corrections as \cite{2024arXiv240712779W} to the three observations. These corrections included energy calibration (for the first observation only), detector world Coordinate System correction, barycenter correction, aspect-solution corrections, and filtering out bad time intervals (such as those related to solar activity). For Vela PWN, the corrections were done following the procedures outlined in \cite{2022Natur.612..658X}. We performed the analysis using tools from a more recent version of \textit{ixpeobssim} (V31.0.1) \citep{2022SoftX..1901194B}, with response matrix updated to version 13. \textit{ixpeobssim} is a public data analysis and simulation framework developed by the IXPE collaboration. Event selection is performed using the xpselect tool, and the polarization is computed using the xpbin tool, PCUBE and PMAPCUBE methods. 

\section{Results}  \label{sec:results}
IXPE observations allowed mapping of the large scale toroidal magnetic field for both the Crab and Vela PWNe.
\cite{2022Natur.612..658X} and \cite{2023NatAs...7..602B} have shown that the toroidal magnetic field is symmetric with respect to the pulsar spin axis, and the direction of the inferred magnetic field broadly follows the shape of the emission torus. 
In this work, we further provide details regarding the consistency between the magnetic field and the shape of the emission torus in these sources.
We reproduced the polarization maps using a top-hat kernel implemented in the \textit{ixpeobssim}, with results shown in Figure \ref{fig:map}. We employed a kernel of $5\,\mathrm{pixels}\times 5\,\mathrm{pixels}$, with each pixel representing a region of $6^{\prime\prime}\times6^{\prime\prime}$. This kernel, therefore, covers a total region of $30^{\prime\prime}\times30^{\prime\prime}$, matching the angular resolution of IXPE. The smoothed analysis was applied to each $6^{\prime\prime}\times6^{\prime\prime}$ pixel with a step length of 5 pixels.

As shown in Figure \ref{fig:map}, the magnetic field is a closed loop for the Crab nebula, while it is an open curve for the Vela nebula. 
This is similar to their torus structure, with the Crab nebula displaying ring-like shapes and the Vela nebula exhibiting double-arcs.
\cite{2004ApJ...601..479N} proposed a three-dimensional torus model by fitting X-ray data for the Crab and Vela nebula.
We use the estimated geometric parameters to compare with the reconstructed magnetic fields for the Crab and Vela PWNe.

\subsection{Crab PWN} 
Two homo-centric elliptical rings \citep{2004ApJ...601..479N} are adopted to represent the Crab PWN's inner ring and the outer torus, as shown in the left panel of Figure \ref{fig:map}. The center position of the torus is located at right ascension =83.6331$^\circ$, declination =22.0157$^\circ$. The semi-major and semi-minor axes of the outer torus are 60.0$^{\prime\prime}$ and 27.2$^{\prime\prime}$, and of the inner ring are 15.6$^{\prime\prime}$ and 7.5$^{\prime\prime}$, respectively. The position angle of the torus projected in the sky is 
$\Psi$=126.31$^\circ$ (North to East). 

\begin{figure}[ht]
\centering
\includegraphics[width=1.0\columnwidth]{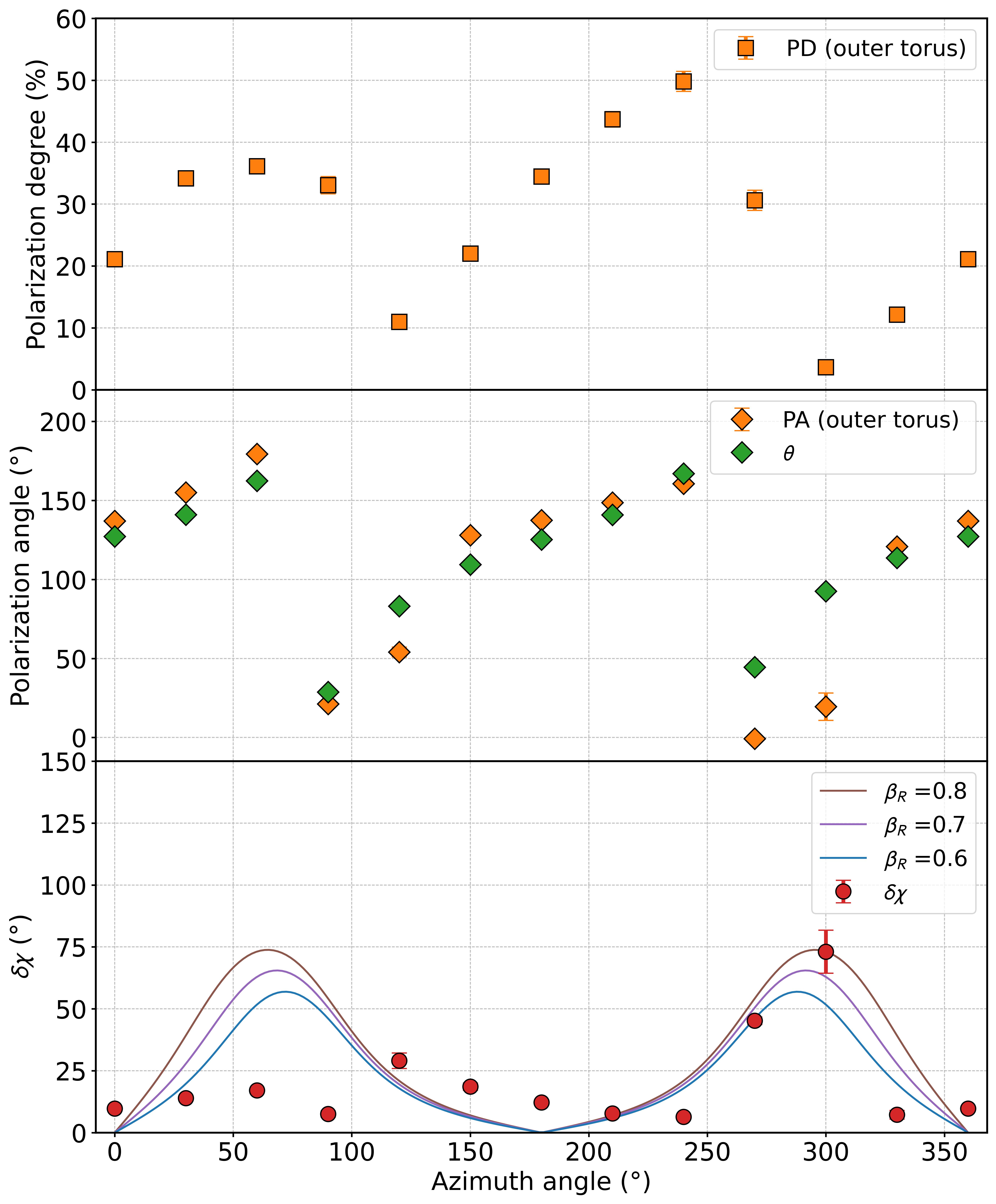}
\caption{The measured PD (top) and PA (center) of regions along the Crab tours as a function of the azimuth angle.
The angle $\theta$ represents the normal line direction of the ellipse along the torus, and the $\delta \chi$ (dots, in bottom) represents the deviation of PA from $\theta$. The PA swing model (solid line, bottom) due to different flow velocities $\beta_{R}=v/c$ is defined in \cite{2017MNRAS.470.4066B} equation (A5).}
\label{fig:crab}
\end{figure}

The inner ring of Crab PWN is too small to be resolved by IXPE. Therefore, we focus on the polarization properties along the outer torus.
We selected 12 independent circular regions with the same radius of 7.5$^{\prime\prime}$ uniformly distributed along the outer torus. We note that compared to a region with a radius of 30$^{\prime\prime}$, a circular region with a radius of 7.5$^{\prime\prime}$ allows for a more focused and precise analysis of polarization within smaller structure of the nebula. In contrast, a larger extraction region of 30$^{\prime\prime}$ could result in depolarization effect. Additionally, we performed simulations in \ref{sec:simulation} to validate the correctness of the smaller circular region. As depicted in the left panel of Figure \ref{fig:image_twonebula}, the white circles and the white points represent these regions and their centers along the outer torus.
The position of the interested region is defined as the azimuth angle of the center counterclockwise from the minor axis of the ellipse in the northwest. Thus the white circular region with red dot represents the region with azimuth angle equals to zero, and the arrow indicates the direction used for selecting the regions.

The corresponding polarization results are shown in Figure \ref{fig:crab}. 
The variation of PD appears to be an asymmetric sinusoidal curve as a function of the azimuth angle. The high PD regions are located in the north ($> 40\%$) and south ($> 50\%$), while the low PD regions are in the north-east and south-west of the torus, where the PA changes rapidly, as shown in the middle panel.
The PA measured along the outer torus exhibits a monotonic trend between 0--180$^\circ$, which is highly consistent with the normal line of the elliptical ring at that center point, represented as $\theta$ in the middle panel of Figure \ref{fig:crab}. The difference between measured PA and $\theta$, annotated as $\delta \chi$, is presented in the bottom panel.

In the north-east and south-west regions of the Crab PWN, the variation of PA is larger than the variation in the geometric structure. According to the model proposed by \cite{2005A&A...443..519B}, the measured $\delta \chi$ can be used to estimate the bulk speeds in the nebulae. As shown in the bottom panel of Figure \ref{fig:crab}, the model predicts the location of the first peak at an azimuth angle between 60$^\circ$ and 75$^\circ$, with maximum values of $\delta \chi$ estimated at approximately 74$^\circ$, 65$^\circ$, and 57$^\circ$, for assumed bulk speeds of 0.8$c$, 0.7$c$, and 0.6$c$, respectively. Both the model and observational results exhibit similar variations in $\delta \chi$ at the position of the second peak, corresponding to the azimuth angle around 280$^\circ$--300$^\circ$. Specifically, the maximum $\delta \chi$ = 72$\pm$10$^\circ$ with significance of 3.0$\sigma$ is consistent with the $\beta$=0.8 model within one error range.

\begin{figure}[ht]
\centering
\includegraphics[width=1.0\columnwidth]
{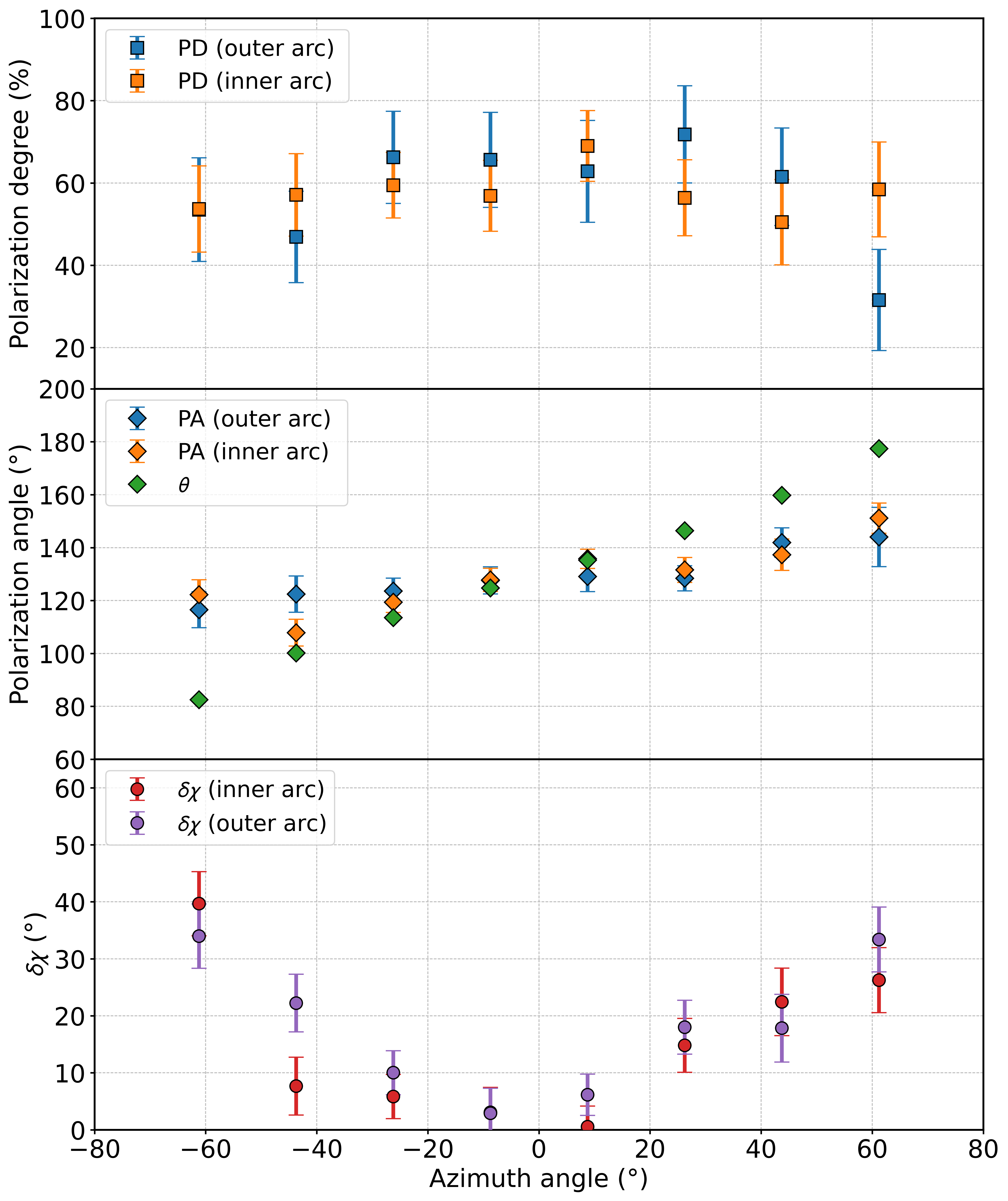}
\caption{The measured PD (top) and PA (center) of regions along the Vela elliptical arcs as a function of the azimuth angle. The definitions of $\theta$ and $\delta \chi$ are the same as Figure \ref{fig:crab}. These results are obtained from a joint analysis of three detector units in the 2--8 keV energy band. The measured PD values are significantly higher than the minimum detectable polarization at the $99\%$ confidence level (MDP$_{99}$).}
\label{fig:arcpolarization}
\end{figure}

\subsection{Vela PWN}

Two parallel elliptical arcs \citep{2004ApJ...601..479N} are used to model the double-arcs shown in Vela PWN X-ray image, as depicted in the right panel of  Figure \ref{fig:map}. 

The center positions of the inner elliptical arc is located at right ascension=128.8398$^\circ$ and declination=-45.1786$^\circ$, and of the outer arc is at right ascension=128.8352$^\circ$ and declination=-45.1766$^\circ$. They have the same semi-major axis of 21.25$^{\prime\prime}$ and semi-minor axis of 15.39$^{\prime\prime}$. The projected separation between the two arcs is 11.61$^{\prime\prime}$. The position angle of the symmetry axis of the torus (semi-minor axis of the ellipse) projected onto the sky is $\Psi$=130$^\circ$. 

We applied the same method used for the Crab PWN to the data analysis of Vela PWN. As depicted in the right panel of Figure \ref{fig:image_twonebula}, we performed polarimetry analysis on eight circular regions with a radius of 5$^{\prime\prime}$, along the double-arcs. The azimuth angles of the interested regions are from -61$^\circ$ to 61$^\circ$, and the results are presented in Figure \ref{fig:arcpolarization}. It is worth mentioning that similar results are obtained when the radius of the circular regions ranges from 5$^{\prime\prime}$ to 15$^{\prime\prime}$. 

The PD of the outer arc fluctuates between approximately 30\% and 70\%, while the PD of the inner arc fluctuates between approximately 50\% and 65\%. This difference could be attributed to the dilution effect caused by the Vela pulsar. The closer proximity of the inner arc to the pulsar results in a greater mixing of emission from the pulsar with the polarized light from the nebula, thereby reducing the observed PD in the inner arc. We conducted the off-pulse analysis according to the off-pulse phase range defined by \cite{1999ApJ...524..373S}. Off-pulse analysis result indicates that the polarization degree along the inner arc is 5\%–10\% higher than integral results. PA derived from the double-arcs regions have similar and small variations, changes from 110$^\circ$ to 150$^\circ$ for the inner arc and from 115$^\circ$ to 140$^\circ$ for the outer arc, respectively. This is quite different from the case of the Crab PWN. The variation in PA is smaller than the variation in the geometric structure for the Vela double-arcs. The PA variation along the azimuth angle is relatively stable compared to the variation in geometric structure. The most considerable deviations, represented by $\delta\chi$, occur at the edges of the double arc, corresponding to the northeast and southwest regions of the PWN, reaching approximately 40$^\circ$. The PA along the inner arc increases at a slightly faster rate than that of the outer arc, which is consistent with the overall magnetic field map of the Vela PWN, that the curvature in the outer region is slightly larger.

\subsection{Simulation}
\label{sec:simulation}

\begin{figure*}[ht]
\centering
\includegraphics[width=\textwidth]{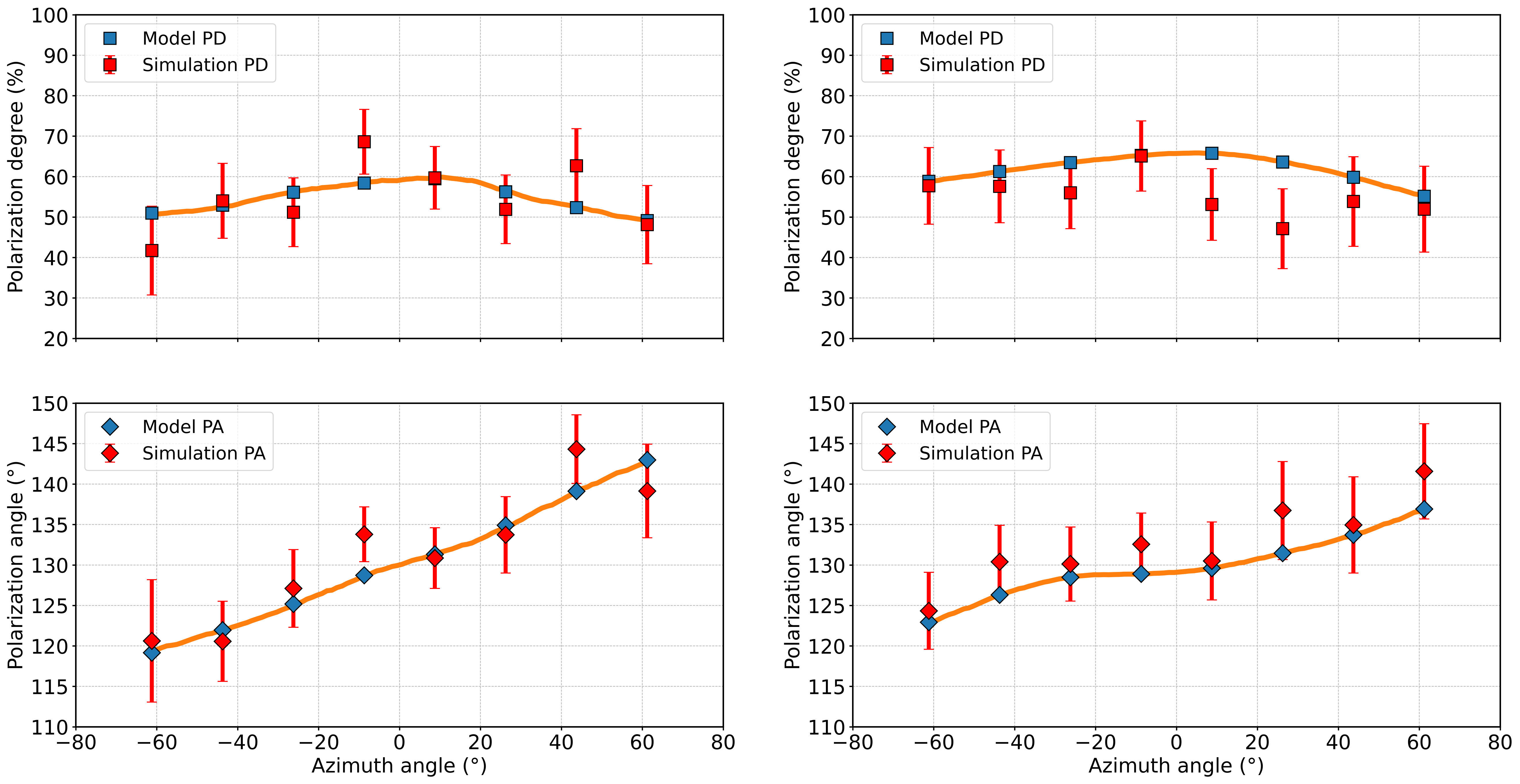}
\caption{Comparison of the simulated PD (top) and PA (bottom) with the input model for Vela PWN inner arc (left) and outer arc (right), as a function of the azimuth angle of the geometry ellipse. The blue model data and the red simulated data are calculated from a round region with a radius of 5$^{\prime\prime}$, while the solid line represents the polarization properties of the model obtained within a circular region with a radius of 2.5$^{\prime\prime}$.} 
\label{fig:vela_arc_sim}
\end{figure*}

There are unavoidable overlaps between adjacent regions in the sliding circular region method, as well as regions smaller than the angular resolution used in the Vela PWN analysis, thus, we employ a simulation with \textit{ixpeobssim} to verify the method. 

The input model for the simulation includes a flux distribution based on the \textit{Chandra} high-resolution image (ObsID: 2815) in the 2--8 keV energy band and a polarization model based on the IXPE observational results described in the right panel of Figure \ref{fig:map}. Both the flux distribution and polarization distribution in the model are represented using a spatial grid of 240$\times$240 with a resolution of 0.492$^{\prime\prime}$.
We conducted the simulation with an exposure time of 860 ks, which is equivalent to the observation duration of IXPE. The same sliding circular region method was applied to the simulated data and the model; results are shown in Figure \ref{fig:vela_arc_sim}. Both the model and the simulated data exhibit a similar trend in PD and PA with respect to the azimuth angle. The maximum deviation in PD is $\sim$15\% and in PA is 5$^\circ$.
The simulated data were obtained by convolving the input model with the IXPE responses, including the effective area, energy resolution, spatial resolution, and modulation response. The differences between the model and the simulation are mainly attributed to the approximately $30^{\prime\prime}$ angular resolution. Fluctuations in the simulated PD and PA for the Vela double-arcs remain within a 2-sigma deviation from  values of input model. This indicates that the method used in this work is reliable, the deviations we found in Vela PWN are intrinsic. 

\section{DISCUSSION AND CONCLUSIONS}
\label{sec:discussion}
We investigated the polarization properties specifically for the Crab outer torus and Vela double-arcs in the energy range of 2--8\,keV. Our current research focuses on these two typical PWNe, which have established simple geometric models. MSH 15-52, another ``torus-jet'' nebula, is not included in this work, and its polarization characteristics have already been reported by \cite{2023ApJ...957...23R}. We plan to extend this analysis to more complex systems, such as MSH 15-52.

The asymmetric PD distribution of the Crab torus implies the nonuniform development of the magnetic field turbulence. The Crab nebula exhibits a complete toroidal magnetic field structure that is closely aligned with the emission geometry of its outer torus. The variation between the polarization vector and the outer torus curvature increases as the nebula moves away from the symmetry axis of the nebula. This phenomenon can be explained by the relativistic effect (PA swing) due to the high velocities of the bulk flow \citep{2005A&A...443..519B}. The maximum $\delta \chi$ of the Crab outer torus reaches 72$\pm$10$^\circ$ with 3.0$\sigma$ significance, suggesting a bulk flow speed of $\sim$0.8$c$ in south-west regions of the Crab PWN.

Figure \ref{fig:vela_arc_sim} illustrates the polarization results obtained from simulations conducted using a toroidal magnetic field consistent with the observations of Vela PWN. These results indicate that the sliding circular region method is able to obtain the intrinsic polarization results. 
It is intriguing that the magnetic field of the Vela PWN does not exhibit the expected alignment with the double-arcs geometry. There is a slight rotation of the PA along the double-arcs, but the curvature of the magnetic field is smaller than that of the emission torus. This finding is inconsistent with most of the magnetohydrodynamic simulations, which usually suggest that the magnetic field vector within the torus should be predominantly aligned with its azimuth \citep{2005A&A...443..519B,2007MNRAS.381.1489N,2017MNRAS.470.4066B}.

The variation in PA along the outer torus of the Crab nebula is larger than that along the double-arcs of the Vela nebula. The disparity in this phenomenon is likely attributable to the distinct physical size between the Crab nebula and the Vela nebula \citep{2001ApJ...556..380H}. The significantly larger Crab nebula exhibits PA variations ranging from 0 to 180$^\circ$, while the smaller Vela nebula shows smaller variations in PA. 

This work is based on the data observed by IXPE, with an angular resolution of 30$^{\prime\prime}$. We clearly see that the PSF of IXPE telescope is affecting the results. Future, more sensitive space-resolved X-ray polarimetry observations of PWNe would provide a deeper understanding of the relationship between X-ray emission structure and the magnetic field.

\section*{Acknowledgments}
This work is supported by National Key R\&D Program of China (grant No. 2023YFE0117200), and National Natural Science Foundation of China (grant No. 12373041), and special funding for Guangxi distinguished professors (Bagui Xuezhe). This work is also supported by the Guangxi Talent Program (“Highland of Innovation Talents”).
This research used data products provided by the IXPE Team (MSFC, SSDC, INAF, and INFN) and distributed with additional software tools by the High-Energy Astrophysics Science Archive Research Center (HEASARC), at NASA Goddard Space Flight Center (GSFC). The Imaging X-ray Polarimetry Explorer (IXPE) is a joint US and Italian mission.  
FLM and ADM contribution is supported by the Italian Space Agency (Agenzia Spaziale Italiana, ASI) through contract ASI-INAF-2022-19-HH.0, and by the Istituto Nazionale di Astrofisica (INAF) in Italy. 
FLM and ADM are partially supported by MAECI with grant CN24GR08 “GRBAXP: Guangxi-Rome Bilateral Agreement for X-ray Polarimetry in Astrophysics”.

\bibliography{reference}
\bibliographystyle{aasjournal}
\end{document}